\documentclass[aps,pra,floatfix,amsmath,amssymb,superscriptaddress]{revtex4}
\usepackage{epsfig}
\usepackage{graphicx}
\usepackage{dsfont}
\begin{document}

\title{Exact Minimum Eigenvalue Distribution of an Entangled Random Pure State}
\author{ Satya N. Majumdar}
\author{Oriol Bohigas}
\affiliation{Laboratoire de Physique Th\'eorique et Mod\`eles Statistiques (UMR 8626 du CNRS), 
Universit\'e Paris-Sud, B\^atiment 100, 91405 Orsay Cedex, France.}
\author{ Arul Lakshminarayan \footnote{Permanent address: Department of Physics, 
Indian Institute of Technology Madras, Chennai, 600036, India.}}
\affiliation{Max-Planck-Institut f\"ur Physik komplexer Systeme, N\"othnitzer 
Stra$\beta$e 38, D-01187 Dresden, Germany.}
\preprint{IITM/PH/TH/2007/13}

\begin{abstract} 

A recent conjecture regarding the average of the minimum eigenvalue of the reduced density 
matrix of a random complex state is proved. In fact, the full distribution of the minimum 
eigenvalue is derived exactly for both the cases of a random real and a random complex state. 
Our results are 
relevant to the entanglement properties of eigenvectors of the orthogonal and unitary 
ensembles of random matrix theory and quantum chaotic systems. 
They also provide a rare exactly solvable case for the distribution
of the minimum of a set of $N$ {\em strongly correlated} random variables for all
values of $N$ (and not just for large $N$).


\end{abstract} 

\maketitle





\newcommand{\newc}{\newcommand}
\newc{\beq}{\begin{equation}}
\newc{\eeq}{\end{equation}}
\newc{\kt}{\rangle}
\newc{\br}{\langle}
\newc{\beqa}{\begin{eqnarray}}
\newc{\eeqa}{\end{eqnarray}}
\newc{\pr}{\prime}
\newc{\longra}{\longrightarrow}
\newc{\ot}{\otimes}
\newc{\rarrow}{\rightarrow}
\newc{\h}{\hat}
\newc{\bom}{\boldmath}
\newc{\btd}{\bigtriangledown}
\newc{\al}{\alpha}
\newc{\be}{\beta}
\newc{\ld}{\lambda}
\newc{\ldmin}{\lambda_{\rm min}}
\newc{\sg}{\sigma}
\newc{\p}{\psi}
\newc{\eps}{\epsilon}
\newc{\om}{\omega}
\newc{\mb}{\mbox}
\newc{\tm}{\times}
\newc{\hu}{\hat{u}}
\newc{\hv}{\hat{v}}
\newc{\hk}{\hat{K}}
\newc{\ra}{\rightarrow}
\newc{\non}{\nonumber}
\newc{\ul}{\underline}
\newc{\hs}{\hspace}
\newc{\longla}{\longleftarrow}
\newc{\ts}{\textstyle}
\newc{\f}{\frac}
\newc{\df}{\dfrac}
\newc{\ovl}{\overline}
\newc{\bc}{\begin{center}}
\newc{\ec}{\end{center}}
\newc{\dg}{\dagger}
\newc{\prh}{\mbox{PR}_H}
\newc{\prq}{\mbox{PR}_q}

\section{Introduction} 

Entangelement has been studied extensively in the recent past due to its central role in 
quantum information and possible involvement in quantum computation 
\cite{NeilsenBook,PeresBook}. It is desirable in many instances to create states of large 
entanglement. Measures of entanglement have been studied mostly in the context of pure bipartite states, 
where the von-Neumann entropy of either subsystem is one of the measures of entanglement 
\cite{PeresBook}.  However there exist other measures of entanglement as well, e.g. the so
called concurrence for two-qubit systems~\cite{Wootters}. 
The entanglement of {\it random pure} quantum states is of interest as they have near 
maximal entanglement content, especially in the context of bipartite entanglement \cite{Winter}. 
Apart from the issue of bipartite entanglement, statistical properties of such random states 
are relevant for quantum chaotic or non-integrable systems.  
The applicability of random matrix theory and hence of random states to systems with well-defined 
chaotic classical limits was pointed out
long back \cite{Bohigas84}. They are also of relevance to other systems with no apparent 
classical limit~\cite{Gaspard,Kudo,Arul}. 

In this paper, we focus on a bipartite quantum system.
More precisely, we consider a bipartite partition of a $NM$-dimensional Hilbert space ${\cal 
H}^{(NM)}$ as ${\cal H}^{(NM)}={\cal H}^{(N)}_A \otimes {\cal H}^{(M)}_B$. We can assume 
without loss of 
generality $N\le M$.
As an example of such a bipartite system, $\cal A$ may be considered a given
subsystem (say a set of spins) and $\cal B$ may represent the environment (e.g., a 
heat bath).
Any quantum state $|\psi\kt$ of the composite system can be generally written as a
linear combination, $|\psi\kt= \sum_{i=1}^N\sum_{\alpha=1}^M x_{i,\alpha}\, 
|i^A\kt\otimes |\alpha^B\kt$ where $|i^A\kt$ and $|\alpha^B\kt$ denote two
complete basis of ${\cal H}^{(N)}_A $ and ${\cal H}^{(M)}_B$ respectively and
the coefficients $x_{i,\alpha}$'s form the entries of a rectangular $(N\times M)$ matrix $X$.
Mutually nonexclusive properties of such a state are entanglement, randomness and statisical
purity. Such a quantum state $|\psi\kt$ is:

\vskip 0.2cm

$\bullet$ {\bf entangled:} if {\it not} expressible as a direct 
product
of two states belonging to the two subsystems $\cal A$ and $\cal B$.  
Only in the special case when the coefficients have the product form,
$x_{i,\alpha}= a_i b_{\alpha}$ for all $i$ and $\alpha$, the state $|\psi \kt= 
|\phi^A\kt \otimes |\phi^B \kt $ can be written as a direct product of two states 
$|\phi^A \kt=\sum_{i=1}^N a_i |i^A\kt$ and 
$|\phi^B\kt= \sum_{\alpha=1}^M b_\alpha |\alpha^B\kt$
belonging respectively to the two subsystems $\cal A$ and $\cal B$. In this case, the
composite state $|\psi \kt$ is {\it fully unentangled}. But otherwise, it is
generically {\it entangled}.

\vskip 0.2cm

$\bullet$ {\bf random:} if the coefficients
$x_{i,\alpha}$ are random variables drawn from an underlyting probability distribution. 
The simplest and the most common random state corresponds to choosing $x_{i,\alpha}$'s
as independent and identically distributed Gaussian variables, real or complex.    

\vskip 0.2cm

$\bullet$ {\bf pure:} if the density matrix of the composite system is simply given by, $\rho=|\psi\kt \br 
\psi|$ with the constraint ${\rm Tr}[\rho]=1$, or equivalently $\br \psi|\psi\kt=1$. 

\vskip 0.2cm

Given a {\it random, pure} and generically {\it entangled} composite state, important
informations on the results of the measurement of any observable on the subsystem $\cal A$
can be derived from the reduced density matrix $\rho_A= {\rm Tr}_B[\rho]$, obtained
upon tracing out the environmental degrees of freedom (i.e., those of subsystem $\cal B$).
It is easy to show (see section II for details) that for a random pure state,
$\rho_A= X X^{\dagger}$ is an $N\times N$ square matrix where $X$ is the $N\times M$
rectangular coefficient matrix. The $N$ unordered eigenvalues $\lambda_1,\lambda_2,\ldots, \lambda_N$
of $\rho_A$ carry important informations regarding the degree of the entanglement
in the subsystem $A$. Given that the entries $x_{i,\alpha}$ of the coefficient matrix $X$
are independent Gaussian variables (real or complex), the eigenvalues $\lambda_i$'s of the 
matrix
$\rho_A= X X^{\dagger}$ are also random variables and their joint probability
density function (jpdf) is known~\cite{LP,ZS}
\beq
P(\ld_1,\ld_2,\cdots,\ld_N)=B_{M,N} \delta\left(\sum_{i=1}^N \ld_i -1 \right)
\prod_{i=1}^{N} \ld_i^{\f{\beta}{2}(M-N+1)-1} \prod_{j<k} |\ld_j-\ld_k|^\beta.
\label{jpdf1}
\eeq
Here $\beta=1,2$ corresponding to the real and complex entries of $A$ and $B_{M,N}$
is the normalization constant that is known explicitly~\cite{ZS}. Several spectral
properties associated with the jpdf in Eq. (\ref{jpdf1}), in particular
for the complex $\beta=2$ case, have been studied
extensively in the literature, for instance see the book ~\cite{ZyckBook} and
references therein.  

In principle, all informations about the spectral properties of the subsystem $\cal 
A$, including its degree of entanglement, are encoded in the jpdf (\ref{jpdf1}).
For example, one useful measure of entanglement is the von Neumann entropy
$S= -\sum_{i=1}^N \lambda_i \ln (\lambda_i)$ which is a random variable.
The average entropy $\br S\kt $ (where the average is performed with the
measure in Eq. (\ref{jpdf1})) was computed for $\beta=2$ by Page~\cite{Page95} and was
found to be $\br S \kt \approx \ln (N)-\frac{N}{2M}$ for large $1<< N\le M$. Noting that
$\ln (N)$ is the maximal possible value of entropy of the subsystem $\cal A$, it follows
that in the limit when $M>>N$, the average entropy, and hence the average
entanglement, of a random pure 
state is
near maximal. Later, the same result was shown to hold for the $\beta=1$ case~\cite{Arul1}. 

While the average entropy is a useful measure of entanglement, it is not 
the unique one. In fact, important informations regarding the nature of
entanglement of a random pure state can also be 
obtained
(see Section II for a detailed discussion)
by studying the probability distributions of the extreme eigenvalues
$\lambda_{\rm max}={\rm max}(\lambda_1,\lambda_2,\ldots, \lambda_N)$
and $\ldmin ={\rm min} (\lambda_1,\lambda_2,\ldots, \lambda_N)$.
In particular, the probability distribution of the minimum eigenvalue
$\ldmin$ provides, in addition to the nature of the entanglement, an important 
information about the degree to which
the effective dimension of the Hilbert space of the subsystem $\cal A$ can be reduced.

In fact, the average value $\br \ldmin\kt$ (with respect to the measure in Eq. (\ref{jpdf1}))
of the minimum eigenvalue was studied recently by Znidaric~\cite{Znd} for the
case $N=M$ and based on 
the exact $\br \ldmin\kt$ for small values of $N$, Znidaric conjectured
that $\br \ldmin\kt=1/N^3$ for all $N$ for the complex case ($\beta=2$). The purpose of this 
paper is to provide {\it exact results for the full probability distribution} of $\ldmin$
for {\it all} $N$ (for the case when $N=M$), 
both
for the complex $(\beta=2$) and the real ($\beta=1$) cases. A byproduct of our
general results is the proof of Znidaric's conjecture for $\beta=2$.
Our results are summarized as 
follows. Let $P_N(x)dx$ denote the probability that $x\le \ldmin\le x+dx$, i.e.,
$P_N(x)$ is the probability density function (pdf) of $\ldmin$. We show that

\vskip 0.2cm

$\bullet$ {\bf Complex case ($\beta=2$):} 

\begin{equation}
P_N(x)= N\,(N^2-1)\, (1-Nx)^{N^2-2}\, \Theta(1-Nx)
\label{beta2}
\end{equation}
where $\Theta(x)$ is the standard Heaviside function, $\Theta(x)=1$ for $x>0$
and $\Theta(x)=0$ for $x<0$. The $k$-th moment $\mu_k(N)=\br \ldmin^k\kt$ is given by
\beq
\mu_k(N)= \frac{\Gamma(k+1)\Gamma(N^2)}{N^k\, \Gamma(N^2+k)}.
\label{umoments1}
\eeq
In particular, for $k=1$, we get $\mu_1(N)= 1/N^3$ thus proving the recent conjecture in 
\cite{Znd}.

\vskip 0.2cm
$\bullet$ {\bf Real case ($\beta=1$):} the result for the real case turns out to be a bit
more complicated. For the pdf of $\ldmin$ we get

\beq
\label{beta1}
P_N(x)=A_N\, x^{-N/2}\, (1-Nx)^{(N^2+N-4)/2} \, _2F_1\left(\f{N+2}{2},\f{N-1}{2},
\f{N^2+N-2}{2},-\f{1-Nx}{x} \right), \; 0 < x \le 1/N
\eeq
and $P_N(x)=0$ for $x \ge1/N$. The constant $A_N$ is given by
\beq
A_N=\df{N\, \Gamma(N)\, \Gamma(N^2/2)}{2^{N-1}\,\Gamma(N/2)\, \Gamma((N^2+N-2)/2)},
\label{AN1}
\eeq
and $_2F_1(\alpha,\beta,\gamma,z)$ is the standard Hypergeometric function defined 
as~\cite{GR}
\begin{equation}
_2F_1(\alpha,\beta,\gamma,z)= 1+ \frac{\alpha\,\beta}{\gamma}\,z + 
\frac{\alpha(\alpha+1)\beta(\beta+1)}{\gamma(\gamma+1)}\,\frac{z^2}{2!}+ 
\frac{\alpha(\alpha+1)(\alpha+2)\beta(\beta+1)
(\beta+2)}{\gamma(\gamma+1)(\gamma+2)}\,\frac{z^3}{3!}+\ldots
\label{hypergeodef}
\end{equation} 
The moments $\mu_k(N)= \br \ldmin^k \kt$ are also computed exactly and are given in Eq. 
(\ref{cmom2}). In particular, the average value ($k=1$) decays for large $N$ as
\begin{equation}
\mu_1(N)\approx \frac{c}{N^3}
\label{avgbeta1}
\eeq
where the prefactor $c$ has a nontrivial value
\beq
c= 2\left[1-\sqrt{\frac{\pi e}{2}}\,
{\rm erfc}(1/\sqrt{2})\right]=0.688641\cdots
\label{constantc}
\eeq

The paper is organized as follows. In Section II, we provide a general introduction to the random
pure states of a bipartite system and recapitulate some general facts leading to the
jpdf (\ref{jpdf1}). Section II and III provide the detailed calculations
of the distribution of the minimum eigenvalue for the complex and the real cases
respectively. Finally we conclude in Section IV with a summary and open questions.
Some details of the calculations are presented in the two appendices.

\section{A Random pure state of a bipartite system}

In this section we 
recall some general facts about a {\it random pure} (RP) state of a bipartite system, its entanglement
properties and the associated random matrix ensemble. As mentioned in the introduction,
let us consider a composite
bipartite system $A\otimes B$ composed of two smaller subsystems $A$ and $B$, whose
respective Hilbert spaces ${\cal H}^{(N)}_A$ and ${\cal H}^{(M)}_B$ have dimensions
$N$ and $M$. The Hilbert space of the composite system ${\cal
H}^{(NM)}={\cal H}^{(N)}_A \otimes {\cal H}^{(M)}_B$ is thus $NM$-dimensional.
Without
loss of generality we will assume that $N\le M$. Let $\{|i^A\kt \}$ and
$\{|\alpha^B \kt \}$ represent two complete basis states for $A$ and $B$
respectively. Then, any arbitrary state $|\psi \kt$ of the composite system
can be most generally written as a linear combination
\beq
|\psi\kt = \sum_{i=1}^N\sum_{\alpha=1}^M x_{i,\alpha}\, |i^A\kt\otimes |\alpha^B\kt
\label{state1}
\eeq
where the coefficients $x_{i,\alpha}$'s form the entries of a rectangular $(N\times M)$ matrix
$X=[x_{i,\alpha}]$. 

Now, the state $|\psi\kt$ is a statistically {\it pure}
state of the composite system if the density matrix of the composite system
is given by
\beq
\rho = |\psi\kt\, \br\psi|.
\label{dm1}
\eeq
Note that had the composite system been in a statistically {\it mixed} state, its density
matrix would have been of the form
\beq
\rho = \sum_{k} p_k\, |\psi_k\kt\, \br \psi_k|,
\label{dmmixed}
\eeq
where $|\psi_k\kt$'s are the pure states of the composite system and $0\le p_k\le 1$
denotes the probability that the composite system is in the $k$-th pure state, with
$\sum_k p_k=1$. In this paper, we will restrict ourselves to the case when
the composite system is in
a pure state denoted by $|\psi\kt$. Then its density matrix in Eq. (\ref{dm1}),
upon using the decomposition in Eq. (\ref{state1}), can be expressed as
\beq
\rho = \sum_{i,\alpha}\sum_{j,\beta} x_{i,\alpha}\, x_{j,\beta}^*\, |i^A\kt\br j^A|\otimes 
|\alpha^B\kt 
\br\beta^B|,
\label{dem2}
\eeq
where the Roman indices $i$ and $j$ run from $1$ to $N$ and the Greek indices $\alpha$ and 
$\beta$ run from $1$ to $M$. We also assume that the pure state $|\psi\kt$ is normalized to 
unity so that ${\rm Tr}[\rho]=1$. Hence the coefficients $x_{i,\alpha}$'s must
be such that ${\rm Tr}[\rho]=1$.

Given the density matrix of the pure composite state in Eq. (\ref{dem2}), one can
then compute the reduced density matrix of, say, the subsystem $\cal A$ by tracing over
the states of the subsystem $\cal B$
\beq
\rho_A = {\rm Tr}_B[\rho]=\sum_{\alpha=1}^M \br \alpha^B|\rho|\alpha^B\kt.
\label{rdm1}
\eeq
Using the expression in Eq. (\ref{dem2}) one gets
\beq
\rho_A = \sum_{i,j=1}^N \sum_{\alpha=1}^M x_{i,\alpha}\, x_{j,\alpha}^*\, |i^A\kt\br 
j^A|=\sum_{i,j=1}^N W_{ij} |i^A\kt\br j^A|
\label{rdm2}
\eeq
where $W_{ij}$'s are the entries of the $N\times N$ square matrix $W=X X^{\dagger}$.
In a similar way, one can express the reduced density matrix $\rho_B={\rm Tr}_A[\rho]$ of the 
subsystem $B$
in terms of the square $M\times M$ dimensional matrix $W'=X^{\dagger}X$.

Let $\lambda_1,\lambda_2,\ldots,\lambda_N$ denote the $N$ eigenvalues of $W=XX^{\dagger}$.
Note that these eigenvalues are nonnegative, $\lambda_i\ge 0$ for all $i=1,2,\ldots,N$.
Now the matrix $W'=X^{\dagger} X$ has $M\ge N$ eigenvalues. It is easy to prove that $M-N$ of them
are identically $0$ and $N$ nonzero eigenvalues of $W^{\dagger}$ are the same as
those of $W$. Thus, in this diagonal representation, one can express $\rho_A$ as
\beq
\rho_A= \sum_{i=1}^N \lambda_i \, |\ld^A_i\kt\, \br \ld^A_i|
\label{diagA}
\eeq
where $|\ld^A_i \kt$'s are the eigenvectors of $W=XX^{\dagger}$. A similar
representation holds for $\rho_B$. It then follows that one can represent 
the original composite state $|\psi\kt$ in this diagonal representation as
\beq
|\psi\kt = \sum_{i=1}^{N} \sqrt{\ld_i}\, |\ld_i^A\kt \otimes |\ld^B_i \kt
\label{Sch1}
\eeq
where $|\ld^A_i \kt$ and $|\ld^B_i \kt$ represent the normalized eigenvectors (corresponding 
to nonzero eigenvalues) of $W=XX^{\dagger}$ and $W'=X^{\dagger} X$ respectively.
This spectral decomposition in Eq. (\ref{Sch1}) is known as the Schimdt decomposition. The 
normalization
condition $\br \psi|\psi\kt=1$, or equivalently ${\rm Tr}[\rho]=1$, imposes
a constraint on the eigenvalues, $\sum_{i=1}^N \ld_i=1$. 

Note that while each individual state $|\ld_i^A\kt \otimes|\ld^B_i \kt$ in the Schimdt decomposition
in Eq. (\ref{Sch1}) is {\it unentangled}, their linear combination $|\psi\kt$, in general,
is {\it entangled}. This simply means that the composite state $|\psi\kt$ can not, in general, 
be written as a direct product $|\psi\kt= |\phi^A\kt \otimes |\phi^B\kt$ of two states of the 
respective subsystems. The spectral properties of the matrix $W$, i.e., the knowledge
of the eigenvalues $\lambda_1,\lambda_2,\ldots, \lambda_N$, in association
with the Schimdt decomposition in Eq. (\ref{Sch1}), provide useful information
about how entangled a pure state is. For example, as mentioned in the introduction, one useful measure
of the entanglement is the von Neumann entropy, $S=-\sum_{i=1}^N \ld_i \ln (\ld_i)$. 

In addition, the two extreme eigevalues,
the largest $\lambda_{\rm max}={\rm
max}(\lambda_1,\lambda_2,\ldots, \lambda_N)$ and the smallest
$\lambda_{\rm min}={\rm
min}(\lambda_1,\lambda_2,\ldots, \lambda_N)$ also provide useful information
about the entanglement.
Note that due to the constraint $\sum_{i=1}^N \ld_i=1$ and the fact that
all eigenvalues are nonnegative, it follows that $1/N\le \ld_{\rm max}\le 1$
and $0\le \ld_{\rm min} \le 1/N$. 
Consider, for 
instance, the following 
limiting situations. Suppose that the largest eigenvalue $\lambda_{\rm max}={\rm 
max}(\lambda_1,\lambda_2,\ldots, \lambda_N)$ takes its maximum allowed value $1$. Then due to
the constraint $\sum_{i=1}^N \lambda_i=1$ and the fact that $\lambda_i\ge 0$ for all $i$,
it follows that all the rest $(N-1)$ eigenvalues must be identically $0$. In that
case, it follows from Eq. (\ref{Sch1}) that $|\psi\kt$ is fully {\it unentangled}.
On the other hand, if $\lambda_{\rm max}=1/N$ (i.e., it takes its lowest allowed value),
it follows that all the eigenvalues must have the same value, $\lambda_i=1/N$ for all $i$,
again due to the constraint $\sum_{i=1}^N \lambda_i=1$. In this case, one can
show that the pure state $|\psi \kt$ is {\it maximally} entangled, as this state
maximizes the von Neumann entropy $S=\ln (N)$. 

In this paper, we will focus on the smallest eigenvalue $0\le \ldmin\le 1/N$. As in the 
case of the largest eigenvalue above, let us consider the two limiting situations. 
When $\ldmin$ takes its maximal allowed value $\ldmin=1/N$, it follows
again from the constraint $\sum_{i=1}^N \lambda_i=1$ that all the eigenvalues
must have the same value $\lambda_i=1/N$. This will thus make the state
$|\psi \kt$ {\it maximally} entangled. In the opposite case, when $\ldmin=0$ takes its smallest
allowed value, while it does not provide any information on the entanglement
of the state $|\psi\kt$, one sees from the Schmidt decomposition that the
dimension of the effective Hilbert space of the subsystem $A$ gets reduced
from $N$ to $N-1$. Indeed, if $\ldmin$ is very close to zero, one can
effectively ignore the term containing $\ldmin$ in Eq. (\ref{Sch1}) and thus
achieve a reduced Hilbert space, a process called `dimensional reduction' that is
often used in the compression of large data structures in computer vision~\cite{Wilks,Fukunaga,VMB}.
Thus the knowledge of 
$\ldmin$ and in particular its proximity
to its upper and lower limits provide informations on both the entanglement phenomenon
as well as on the efficiency of the dimensional reduction process.  

So far, our discussion is valid for an arbitrary pure state in Eq. (\ref{state1}) with
any fixed coefficient matrix $X=[x_{i,\alpha}]$. Now, such a pure state will
be called a {\it random pure} state if the coefficients $x_{i,\alpha}$'s are 
random variables, drawn from an underlying probability distribution. In particular, we
will consider the case when the elements of $X$ are independent and identically
distributed random variables, real or complex, drawn from a Gaussian distribution:
${\rm Prob}[X]\propto \exp\left[-\frac{\beta}{2} {\rm Tr}(X^{\dagger} X)\right]$, where
the Dyson index $\beta=1,2$ corresponds respectively to the real and complex 
$X$ matrices.  
The product $W=XX^{\dagger}$ is called the random Wishart matrix~\cite{Wishart}.
The joint distribution of the $N$ nonnegative eigenvalues of $W$ is known~\cite{James}
\beq
P^{W}(\lambda_1,\lambda_2,\ldots,\lambda_N)\propto \, e^{-\frac{\beta}{2}\sum_{i=1}^N 
\lambda_i}\, \prod_{i=1}^N \lambda_i^{\frac{\beta}{2}(1+M-N)-1}\, \prod_{j<k} 
|\lambda_j-\lambda_k|^{\beta}.
\label{wishart1}
\eeq

Note however, that in case of a {\it random pure} state $|\psi\kt$ in Eq. (\ref{state1}),
the eigenvalues of the matrix $W=XX^{\dagger}$ are not quite the same as that
of the Wishart matrix, due to the additional constraint that ${\rm Tr}[\rho]={\rm Tr}[W]=1$.
Thus, the eigenvalues of $W$ that appear in the Schimdt decomposition in Eq. (\ref{Sch1}),
are distributed according to the Wishart law in Eq. (\ref{wishart1}), but in addition
have to satisfy the constraint $\sum_{i=1}^N \lambda_i=1$. This constraint can be explicitly incorporated
by multiplying a delta function $\delta(\sum_{i=1}^N \lambda_i-1)$ to the Wishart measure in 
Eq. (\ref{wishart1}). With this additional delta function multiplying the Wishart measure, 
the exponential term in Eq. (\ref{wishart1}) 
just becomes a constant
and can be absorbed into the overall normalization constant and one arrives at the
jpdf of the eigenvalues of $W$ mentioned in Eq (\ref{jpdf1}) in the introduction.

Given the jpdf (\ref{jpdf1}), we are interested here in the distribution 
of the minimum eigenvalue $\ldmin$. Let $Q_{N,M}(x)={\rm Prob}[\ldmin\ge x]$ be the
cumulative distribution of $\ldmin$. The pdf of $\ldmin$ is simply obtained
by taking the derivative, $P_{N,M}(x)= -dQ_{N,M}(x)/dx$. Since the event $\ldmin\ge x$
necessarily implies that all the eigenvalues $\lambda_i\ge x$ (for all $i=1,2,\ldots,N$), 
it follows, upon using the explicit jpdf (\ref{jpdf1}),
that $Q_{N,M}(x)$ is precisely given by the multiple integral (with $N\le M$)
\beq
Q_{N,M}(x)= B_{M,N}\, \int_x^{\infty}\cdots \int_x^{\infty}  \delta\left(\sum_{i=1}^N 
\ld_i -1 \right)
\prod_{j<k} |\ld_j-\ld_k|^\beta\, \prod_{i=1}^N 
\ld_i^{\f{\beta}{2}(M-N+1)-1}\, d\lambda_i.
\label{qmin}
\eeq
The real technical challenge is to evaluate this multiple integral. In the next two
sections, we show
how to compute this integral exactly respectively for $\beta=2$ and $\beta=1$,
for all $M=N$, i.e.,  when 
the Hilbert spaces
of the two subsystems have equal dimensions. In this case, i.e., when $M=N$, we
will denote, for simplicity of notations, $Q_{N,N}(x)=Q_N(x)$ for
the cumulative distribution of the minimum eigenvalue and the corresponding
density by $P_{N,N}(x)=P_N(x)=-dQ_N(x)/dx$.

\section{A Complex Random Vector} 

This section is devoted to 
finding exactly the distribution of the minimum 
eigenvalue $\ldmin$ or the minimum Schmidt coefficient for random complex states.
Let 
\beq
Q_N(x)=\mbox{Prob}\left[ \ldmin \ge x \right]=
\mbox{Prob}\left[ \ld_{1} \ge x, \ld_2 \ge x, \ldots, \ld_N \ge x \right].
\eeq
Therefore
\beq
Q_N(x)=B_{N,N} \int_{x}^{\infty} \cdots 
\int_{x}^{\infty} \delta\left(\sum_{i=1}^N \ld_i -1 \right) \prod_{j<k} (\ld_j-\ld_k)^2
\prod_{i=1}^N \, d \ld_i
\eeq
An evaluation of this multiple integral proceeds by introducing an auxiliary one defined by
\beq
I(x,t)=\int_{x}^{\infty} \cdots \int_{x}^{\infty} 
\delta\left(\sum_{i=1}^N \ld_i -t \right) \prod_{j<k} (\ld_j-\ld_k)^2
\prod_{i=1}^N \, d \ld_i,
\eeq
so that $Q_N(x)=B_{N,N}\, I(x,1)$. Consider the  following Laplace transform of $I(x,t)$:
\beq
\int_0^{\infty} I(x,t) e^{-st} dt = \int_{x}^{\infty} \cdots 
\int_{x}^{\infty} e^{-s \sum_{i=1}^{N} \ld_i} \prod_{j<k} (\ld_j-\ld_k)^2
\prod_{i=1}^N \, d \ld_i.
\eeq
A linear shift and scaling $z_i=s(\ld_i-x)$ results in 
\beq
\int_0^{\infty} I(x,t) e^{-st} dt = \df{e^{-sNx}}{s^{N^2}} \int_{0}^{\infty} 
\cdots \int_{0}^{\infty}e^{-\sum_{i=1}^{N} z_i} \prod_{j<k} (z_j-z_k)^2
\prod_{i=1}^N \, d z_i .
\eeq
Thus the dependence on $s$ and $x$ just factors out of the integral. The integral
happens to be one of the Selberg integrals which can be evaluated
explicitly~\cite{Mehta} and this gives 
\beq
\int_0^{\infty} I(x,t) e^{-st} dt = \df{e^{-sNx}}{s^{N^2}} \prod_{j=0}^{N-1} \Gamma(j+2) \Gamma(j+1).
\eeq
An inverse Laplace transform yields
\beq
I(x,t)=\frac{\prod_{j=0}^{N-1} \Gamma(j+2) \Gamma(j+1)}{ \Gamma(N^2)}\, 
\left(t-Nx\right)^{N^2-1}\, 
\Theta\left(t-Nx \right).
\eeq
Using the known normalization constant~\cite{ZS} 
\beq
B_{N,N}=\frac{\Gamma(N^2)}{  \prod_{j=0}^{N-1} \Gamma(N-j) \Gamma(N-j+1)}
\eeq
we finally arrive at 
\beq
Q_N(x)=\mbox{Prob}\left[ \ldmin \ge x \right] = B_{N,N}\,I(x,1)=\left(1-N x\right)^{N^2-1} 
\Theta\left(1-Nx \right).
\label{udist}
\eeq
Subsequently, the pdf is given by 
\beq
P_N(x)=-\frac{dQ_N(x)}{dx}= N(N^2-1) (1-Nx)^{N^2-2}\,\Theta(1-Nx).
\label{updf}
\eeq
A plot of this pdf can be found in Fig. 1 for $N=4$. Thus $P_N(x)$ in $x\in [0,1/N]$ has the 
limiting behavior 
\begin{eqnarray}
P_N(x) & \to & N\,(N^2-1) \quad\quad {\rm as} \quad x\to 0 \nonumber \\
       & = & N\,(N^2-1)\,(1-Nx)^{N^2-2} \quad {\rm as} \quad x\to 1/N 
\label{clim}
\end{eqnarray}
Note that in the regime where $x<< 1/N$, the pdf in Eq. 
(\ref{updf}) becomes
exponential, $P_N(x) \approx N(N^2-1)\exp[-N(N^2-1)x]$.
Let us also note that the distribution of the smallest eigenvalue in Eq. (\ref{udist}) is identical 
to that of the smallest intensity component of a complex random state
derived recently \cite{ASOS}, provided one replaces $N^2$ (in the exponent in Eq. 
(\ref{udist})) by $N$. 

\vskip 0.2cm

\noindent {\bf Moments of $\ldmin$:} From the explicit expression of the pdf in Eq. 
(\ref{updf}) one can easily compute all
the moments of $\ldmin$. For the $k$-th moment we get
\beq
\mu_k(N)= \br \ldmin^k \kt=\int_0^{\infty} x^k P_N(x)\, 
dx=\frac{\Gamma(k+1)\Gamma(N^2)}{N^k\, \Gamma(N^2+k)}.
\label{umoments}
\eeq
In particular, for $k=1$, we obtain for all $N$
\beq
\mu_1(N)=\br \ldmin \kt=\frac{1}{N^3}, 
\label{uavg}
\eeq
thus proving the recent conjecture by Znidaric~\cite{Znd} based on evaluations for 
small $N$. Putting $k=2$ in Eq. (\ref{umoments}), we get the second moment $\mu_2=\frac{2}{N^4 
(N^2+1)}$.
Thus the variance is given by
\beq
\sigma^2= \mu_2(N)-[\mu_1(N)]^2= \frac{1}{N^6}\left(\frac{N^2-1}{N^2+1}\right).
\label{uvar}
\eeq

\section{A real random vector}

While complex random vectors are ``generic", real vectors are important as well. For instance in the case 
when the system has a time-reversal symmetry or any anti-unitary symmetry the eigenfunctions can be in 
general chosen to be real and the relevant ensembles are the ``orthogonal'' ones (such as the Gaussian
orthogonal ensemble and the circular orthogonal ensemble), wherein general orthogonal transformations leave the 
ensemble invariant \cite{Mehta,HaakeBook}. The 
entanglement properties of real and complex random states may, in general, differ. For instance 
for so 
called ``single-particle'' states or one-magnon states, real states have lower entanglement measured in terms 
of two-spin entanglement content than the case of the complex states \cite{ArulSub}. In general, much less is 
known 
for random real states than the complex ones, although for instance several many-body Hamiltonians (say of spins) 
have natural time-reversal symmetry. In this section the distribution of the minimum eigenvalue of the real 
case is calculated exactly.

The jpdf of the eigenvalues $\ld_i$ in this case (we again restrict ourselves to the case $M=N$)
is
\beq
\label{realjpdf}
P_{N}(\ld_1,\cdots,\ld_N)=C_{N,N}\delta\left(\sum_{i=1}^{N}\ld_i-1 \right) \prod_{j<k}|\ld_j-\ld_k| 
\prod_{i=1}^N
\df{1}{\sqrt{\ld_i}},
\eeq
where $C_{N,N}$ is the normalization constant and is known to be~\cite{ZS} 
\beq
C_{N,N}^{-1}=\df{2^N}{\pi^{N/2} \Gamma(N^2/2)} \prod_{j=0}^{N-1} 
\Gamma\left(\f{j+1}{2}\right) \Gamma\left(\f{j+3}{2}\right).
\label{rnorm}
\eeq
The cumulative distribution of the smallest eigenvalue, $Q_N(x)={\rm Prob}[\ldmin \ge x]$, is given 
by
\beq
Q_N(x)=C_{N,N} \int_{x}^{\infty} \cdots
\int_{x}^{\infty} \delta\left(\sum_{i=1}^N \ld_i -1 \right) \prod_{j<k} |\ld_j-\ld_k| \prod_{i=1}^N
\df{1}{\sqrt{\ld_i}} \, d \ld_i .
\label{rcum}
\eeq

To evaluate this multiple integral, we proceed, as in the previous section, by defining an auxiliary integral 
$J(x,t)$ as 
\beq
\label{Ireal}
J(x,t)=\int_{x}^{\infty} \cdots \int_{x}^{\infty} \delta\left(\sum_{i=1}^N \ld_i -t \right) 
\prod_{j<k} |\ld_j-\ld_k|
\prod_{i=1}^N \f{1}{\sqrt{\ld_i}} \, d \ld_i,
\eeq
so that $Q_N(x)=C_{N,N}\, J(x,1)$.

Taking the Laplace transform of Eq.~(\ref{Ireal}) leads to
\beq
\int_0^{\infty} J(x,t) e^{-st} dt = \df{1}{(2s)^{N^2/2}} \int_{2sx}^{\infty} \cdots 
\int_{2sx}^{\infty} e^{-\f{1}{2} \sum_{i=1}^N y_i} \prod_{j<k} |y_j-y_k|
\prod_{i=1}^N \f{1}{\sqrt{y_i}} \, d y_i,
\eeq
where the scaled variable $y_i=2s \ld_i$.
We next use a result due to Edelman \cite{Edelman} for the Wishart orthogonal ensemble whose jpdf
is given by
\beq
P^W_N(y_1,\cdots,y_N)=a_{N,N}e^{-\f{1}{2}\sum_{i=1}^N y_i } \prod_{i=1}^N \df{1}{\sqrt{y_i}} \prod_{j<k} 
|y_j-y_k|
\eeq
where the normalization constant $a_{N,N}$ is
\beq
a_{N,N}= \df{C_{N,N}}{2^{N^2/2} \Gamma(N^2/2)}.
\eeq
For such an ensemble Edelman \cite{Edelman} showed that the distribution of the smallest eigenvalue
$Q^W(z)=\mbox{Prob} \left[ y_{\rm min}\ge z \right]$ is given explicitly by
\beq 
Q^W(z)= \int_{z}^{\infty} \cdots \int_{z}^{\infty} P^W(y_1,\cdots,y_N) \prod_{i=1}^N 
dy_1\cdots dy_N=\df{N \Gamma(N)}{2^{N-1/2} \Gamma(N/2)} \int_z^{\infty} 
\df{e^{-Ny/2}}{\sqrt{y}}\,  U\left(\df{N-1}{2},-\df{1}{2},\df{y}{2} \right) \, dy.
\eeq
where $U(a,b;z)$ is the confluent hypergeometric function \cite{AS} of the second kind that satisfies the 
differential equation
\beq
z \df{d^2 U}{dz^2} + (b-z) \df{dU}{dz}-a U =0
\eeq 
with the boundary conditions 
\beq
U(a,b,0)=\df{\Gamma(1-b)}{\Gamma(1+a-b)},\;\; U(a,b,z \rarrow \infty) = 0.
\eeq

Working back we therefore obtain
\beq
\label{laptran}
\int_0^{\infty} J(x,t) e^{-st} dt = \df{1}{(2s)^{N^2/2}} 
\left[ \df{N \Gamma(N)}{2^{N-1/2} \Gamma(N/2) a_{N,N}} \right] 
\int_{2sx}^{\infty} \df{e^{-Ny/2}}{\sqrt{y}}\, U\left(\df{N-1}{2},-\df{1}{2},\df{y}{2} \right) 
\, dy.
\eeq
To make further progress, it turns out to be easier to work with the probability density function 
rather than the cumulative distribution $Q_N(x)$, 
\beq
\label{densitydefn}
P_N(x)= -\df{dQ_N(x)}{dx} = -C_{N,N} \frac{d\,J(x,1)}{dx}.
\eeq 
Taking the derivative of Eq.~(\ref{laptran}) with respect to $x$ leads to
\beq
-\int_{0}^{\infty} \df{dJ(x,t)}{dx} e^{-st} dt = b_N\df{1}{\sqrt{x}} 
\df{e^{-Nsx}}{s^{(N^2-1)/2}} U\left(\df{N-1}{2},-\df{1}{2},sx \right),  
\eeq
where 
\beq
b_N=\df{N \Gamma(N)}{\Gamma(N/2) a_{N,N} 2^{(N^2+2N -2)/2}}.
\eeq

The task then is to find the Laplace inverse:
\beq
\label{dervx}
-\df{dJ(x,t)}{dx} = \df{b_N}{\sqrt{x}} {\cal L}_s^{-1} 
\left[ \df{e^{-Nsx}}{s^{(N^2-1)/2}} U\left(\df{N-1}{2},-\df{1}{2},sx \right) \right].
\eeq
First, an application of the convolution theorem leads to 
\beq
{\cal L}_s^{-1} \left[ \df{e^{-Nsx}}{s^{(N^2-1)/2}}\right] = \df{1}{\Gamma\left( \df{N^2-1}{2} \right)}
\left(t-Nx \right)^{(N^2-3)/2} \Theta\left(t-Nx \right).
\label{inv1}
\eeq
Second, using an integral representation of the hypergeometric function $U(a,b,z)$  \cite{AS} namely
\beq
U(a,b,z)=\df{1}{\Gamma(a)} \int_0^{\infty} e^{-zt} t^{a-1} (1+t)^{b-a-1} dt
\eeq
one obtains the following inverse:
\beq
{\cal L}_s^{-1} \left[ U\left(\df{N-1}{2},-\df{1}{2},sx \right) \right] = \df{x^{3/2}}{\Gamma(\f{N-1}{2})} 
t^{(N-3)/2} (x+t)^{-(N+2)/2}.
\label{inv2}
\eeq
Using the two inverses in Eqs. (\ref{inv1}) and (\ref{inv2}) and the convolution theorem,
we get upon simplifying
\beqa
{\cal L}_s^{-1} \left[ \df{e^{-Nsx}}{s^{(N^2-1)/2}} U\left(\df{N-1}{2},-\df{1}{2},sx \right) \right]
&=&\df{x^{3/2}}{\Gamma\left(\f{N^2-1}{2} \right) \Gamma \left( \f{N-1}{2} \right)} \int_0^{t-Nx} 
t'^{\f{N-2}{2}} \left( x+t' \right)^{-\f{N+2}{2}} \left( t-t'-Nx \right)^{\f{N^2-3}{2}} dt' \\
&=& \df{x^{-(N-1)/2}}{\Gamma\left(\f{N^2+N-2}{2}  \right)} (t-Nx)^{\f{N^2+N-4}{2}} \,
 _2F_1\left(\f{N+2}{2},\f{N-1}{2},\f{N^2+N-2}{2},-\f{t-Nx}{x} \right). \nonumber 
 \eeqa
Here $_2F_1(a,b,c,z)$ is the standard hypergeometric function \cite{AS}, and the integral can
be found in \cite{GR}.
Using this along with Eqs.~(\ref{densitydefn},\ref{dervx}) and substituting $t=1$, we finally get the 
p.d.f. of the minimum
eigenvalue $\ldmin$ as 
\beq
\label{resultreal}
P_N(x)=A_N\, x^{-N/2}\, (1-Nx)^{(N^2+N-4)/2} \, _2F_1\left(\f{N+2}{2},\f{N-1}{2},
\f{N^2+N-2}{2},-\f{1-Nx}{x} \right), \; 0 < x \le 1/N
\eeq
and $P_N(x)=0$ for $x \ge1/N$. The constant $A_N$ is given by 
\beq
A_N=\df{N\, \Gamma(N)\, \Gamma(N^2/2)}{2^{N-1}\,\Gamma(N/2)\, \Gamma((N^2+N-2)/2)}.
\label{AN} 
\eeq

This solves exactly for the distribution of the minimum eigenvalue of the reduced
density matrices of bipartite random real states when the dimensions of the subspaces
are equal.  In the simplest possible case of real states of two qubits, $N=2$, the
distribution is simply
\beq
P_2(x)=\df{1-2x}{\sqrt{x(1-x)}}, \; 0 <x \le 1/2; \quad\; P_2(x)=0, \; x \ge 1/2.
\eeq
This follows from Eq.~(\ref{resultreal}) as $_2F_1(2,1/2,2,x)=1/\sqrt{1-x}$. Alternatively it almost 
immediately follows from the jpdf in Eq.~(\ref{realjpdf}) as there are only two eigenvalues
that sum to unity in this case, and the distribution of the one which is less than one-half
is precisely $P_2(x)$. In Fig. 1, we plot the pdf $P_N(x)$ of $\ldmin$ for $N=4$, both for
the complex case given in Eq. (\ref{updf}) and the real case given in Eq. (\ref{resultreal})

In appendix-A, we work out the limiting behavior of $P_N(x)$ as $x\to 0$ and $x\to 1/N$.
For general $N$, one finds
\begin{eqnarray}
P_N(x) &\approx & \left[\frac{\sqrt{\pi}\, 
\Gamma(N)\,\Gamma(N^2/2)}{2^{N-1}\,\Gamma^2(N/2)\,\Gamma((N-1)/2)}\right]\, x^{-1/2}\quad\, 
{\rm 
as} \quad x\to 0 \nonumber \\
&\approx & A_N\, N^{-N/2}\, (1-Nx)^{(N^2+N-4)/2} \quad\, {\rm
as} \quad x\to 1/N
\label{rlim}
\end{eqnarray}
Comparing this limiting behavior in the real case in Eq. (\ref{rlim}) with that of the 
complex case in Eq. (\ref{clim}) one finds that
while in the former $P_N(x)$ diverges as $x^{-1/2}$ as $x\to 0$, in the latter
it approaches a constant. In the other limit $x\to 1/N$, both the densities
approach zero as a power law $(1-Nx)^{\nu}$, but with different exponents
$\nu=N^2-2$ (for the complex case) and $\nu=(N^2+N-4)/2$ for the real case.

\begin{figure}[ht]
\epsfig{file=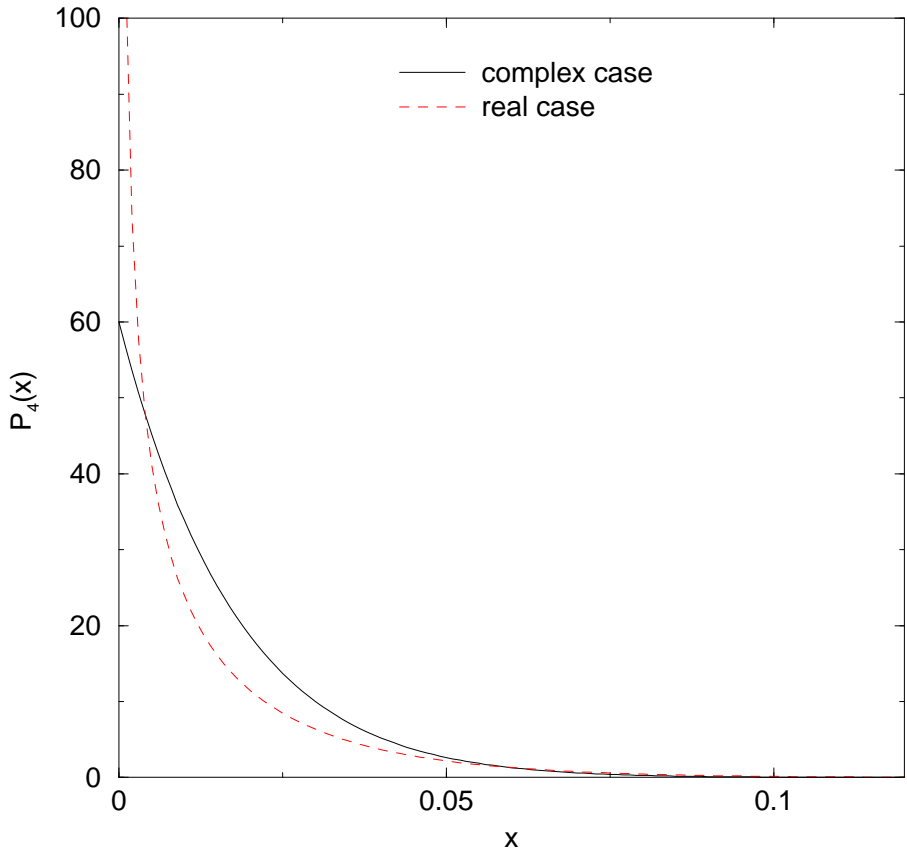,width=12cm} 
\caption{The p.d.f $P_N(x)$ of the minimum eigenvalue $\ldmin$ vs. $x$ for $N=4$, for the
complex and the real cases (Eqs. (\ref{updf}) and (\ref{resultreal}) respecively).
In the complex case, the density approaches a constant
as $x\to 0$, whereas for the real case, it diverges as $x^{-1/2}$ as $x\to 0$.} 
\label{fig:1}
\end{figure}

\vskip 0.2cm

\noindent {\bf Moments of $\ldmin$:} One can use the explicit result for the p.d.f. $P_N(x)$
of $\ldmin$ in Eq. (\ref{resultreal}) to calculate its $k$-th moment
\begin{eqnarray}
\mu_k(N)= \br {\ldmin}^k \kt &=& A_N\, \int_0^{1/N} x^{k-N/2}\, (1-Nx)^{(N^2+N-4)/2}\, 
_2F_1\left(\frac{N+2}{2},\frac{N-1}{2}, \frac{N^2+N-2}{2}, -\frac{1-Nx}{x}\right)\, dx \nonumber \\
&=& A_N \int_0^{\infty} y^{(N^2+N-4)/2}\, (N+y)^{-(N^2+2k)/2}\, _2F_1\left(\frac{N+2}{2},\frac{N-1}{2}, 
\frac{N^2+N-2}{2}, -y\right)
\label{cmom1}
\end{eqnarray}
where we made a change of variable $y=-N+1/x$ in the first line. We next use the following known
integral~\cite{GR}
\beq
\int_0^{\infty} x^{\gamma-1}\, (x+z)^{-\sigma}\, _2F_1(\alpha,\beta,\gamma,-x)\,dx
= 
\frac{\Gamma(\gamma)\Gamma(\alpha-\gamma+\sigma)\Gamma(\beta-\gamma+\sigma)}{\Gamma(\sigma)
\Gamma(\alpha+\beta-\gamma+\sigma)}\, _2F_1(\alpha-\gamma+\sigma, \beta-\gamma+\sigma, 
\alpha+\beta-\gamma+\sigma, 1-z)
\label{iden1}
\eeq
in Eq. (\ref{cmom1}) and also the value of $A_N$ from Eq. (\ref{AN}) to arrive at an explicit 
expression for the $k$-th moment (valid for all $N$),
\beq
\mu_k(N)= \frac{\Gamma(N+1)\Gamma(N^2/2)\Gamma(k+2)\Gamma(k+1/2)}{\Gamma(N/2)\Gamma(k+N^2/2)\Gamma(k+(N+3)/2) 
2^{N-1}}\, _2F_1\left(k+2,k+1/2,k+\frac{N+3}{2}, 1-N\right).
\label{cmom2}
\eeq
One can verify that $\mu_0(N)=1$, thus ensuring the correct normalization. For the average value
of $\ldmin$ we use $k=1$ and get
\beq
\mu_1(N)= \br \ldmin \kt = \frac{\sqrt{\pi}\,\Gamma(N)}{N\, \Gamma(N/2)\,\Gamma((N+5)/2) 
2^{N-1}}\, 
_2F_1\left(3,\frac{3}{2},\frac{N+5}{2},1-N\right).
\label{ravg}
\eeq
Thus the expression for $\br \ldmin \kt$ for arbitrary $N$ in the real case is
considerably more complicated than its counterpart in Eq. (\ref{uavg}) for the complex case.
One finds, from Eq. (\ref{ravg}), that $\mu_1(N)$ decreases with increasing $N$, e.g., $\mu_1(1)=1$, 
$\mu_1(2)=(4-\pi)/8$, $\mu_1(3)=(2-\sqrt{3})/9$ etc. In appendix-B, we show that
asymptotically for large $N$, $\mu_1(N)$ decays as 
\beq
\mu_1(N) \approx \frac{c}{N^3};\quad\, {\rm where}\quad\, c=2\left[1-\sqrt{\frac{\pi e}{2}}\, 
{\rm erfc}(1/\sqrt{2})\right]=0.688641\cdots
\label{ravgas}
\eeq
where ${\rm erfc}(x)=\frac{2}{\sqrt{\pi}}\int_x^{\infty} e^{-u^2}\, du$ is the complementary error function.
The large $N$ result in Eq. (\ref{ravgas}) for the real case should be compared to that of the complex case
where $\mu_1(N)=1/N^3$. One sees that the average value of the minimum eigenvalue in the former case
is less by a constant factor $c=0.688641\cdots$ compared to the later case. 
In Fig. 2, we plot both the exact formula for $\mu_1(N)$ in Eq. (\ref{ravg}) and the asymptotic form
in Eq. (\ref{ravgas}) against $N$.

\begin{figure}[ht]
\epsfig{file=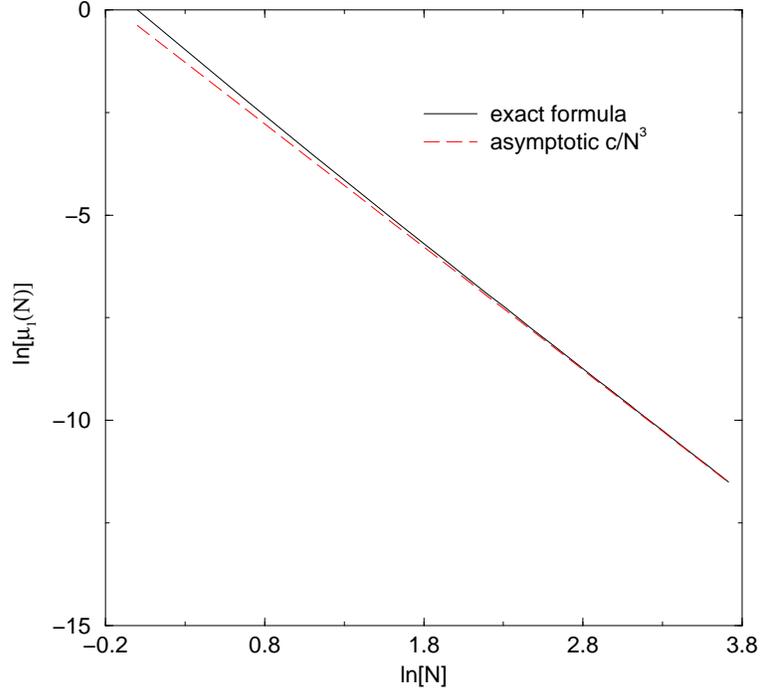,width=10cm}
\caption{A log-log plot of the exact formula of $\mu_1(N)$ in Eq. (\ref{ravg}) vs. $N$ 
compared with
the asymptotic formula $\mu_1(N)\approx c/N^3$ vs. $N$ with $c=0.688641$ for the real case.}
\label{fig:2}
\end{figure}

\section{Conclusion}

In this paper we have computed the exact probability distribution of the minimum eigenvalue
$\ldmin$ of an entangled random (both real and complex) pure state of a bipartite system composed of two 
subsystems whose respective Hilbert spaces have equal dimensions $M=N$. We have also computed exactly
all the moments
of $\ldmin$ for all $N$. As a byproduct, we prove that $\br \ldmin\kt=1/N^3$ for all
$N$ for complex matrices, a result recently conjectured~\cite{Znd}.
The pdf of the minimum eigenvalue in the real case differs significantly from its
complex counterpart. 

Apart from providing important informations on the nature of the entanglement of a random pure state
as well as on the degree to which the dimension of the Hilbert space of a subsystem can be 
reduced,
our result for the distribution of the minimum eigenvalue 
has some relevance in the general context of extreme value statistics. This 
subject has been around for a long time~\cite{GumbelBook}, but has seen a 
recent resurgence 
due to its many applications in diverse areas such as engineering, economics and physical
sciences \cite{KantzBook}.  
If the underlying random variables are {\it independent and identically distributed} then there 
are three possible limiting universal distributions for the 
extreme events, the Fr\'echet, the Gumbel and the 
Weibull distributions. However, much less is known when the underlying random variables are {\it strongly 
correlated}.
In such cases, the limiting distribution (for large $N$) of the maximum is known exactly only in very
few cases. 
For example, the limiting distribution of the largest eigenvalue of 
a $N\times N$ Gaussian unitary random matrix (GUE) is given by the celebrated Tracy-Widom 
law~\cite{TracyWidom}, which has found many recent applications~\cite{leshouches}.
Similarly, the Tracy-Widom law also describes the limiting distribution of
the largest eigenvalue of Wishart matrices~\cite{Johansson,Johnstone}, for random
matrices with certain non-Gaussian entries~\cite{BBP} and the
scaled height of a $(1+1)$-dimensional growth models~\cite{PS,leshouches}. 
The probabilities of large deviations of $\lambda_{\rm max}$, outside
the regime of the validity of the Tracy-Widom law, have also been
computed recently both for Gaussian~\cite{DM} and Wishart matrices~\cite{Johansson,VMB}.  
Other examples for which the limiting distribution is known exactly include 
the maximum relative height of a class of one dimensional fluctuating interfaces in their
steady states in a finite system~\cite{MC,SM} and 
$1/f^{\alpha}$ noise signals~\cite{Racz}.
In contrast, much less is known about the distribution of the extreme eigenvalues for
finite $N$, a notable exception being the minimum eigenvalue
for $N\times N$ Wishart matrices whose distribution was computed exactly by Edelman for all $N$~\cite{Edelman}.  
In our present context, the eigenvalues of a random pure state are also strongly correlated due 
to the
presence of the Vandermonde term $\prod_{j<k} |\lambda_j-\lambda_k|^{\beta}$ in the
jpdf (\ref{jpdf1}). So our results 
provide another rare exactly solvable case
for the distribution of the minimum of a set of $N$ {\it strongly correlated} random variables,
and this is not just for large $N$ but
for {\it any finite} $N$.

Computing the distribution of $\ldmin$ for unequal dimensions ($M\neq N$) of the Hilbert spaces
of the subsystems remains a challenging open problem.
 
\appendix
\section{Limiting behavior of $P_N(x)$ for the real case}

In this appendix, we derive the behavior of $P_N(x)$ in Eq. (\ref{rlim}), starting
from the exact expression of $P_N(x)$ in Eqs. 
(\ref{resultreal}) and (\ref{AN}). The behavior near the upper limit $x\to 1/N$
is simple to derive. Using $_2F_1(\alpha,\beta,\gamma,0)=1$, one immediately finds from
Eq. (\ref{resultreal}) that as $x\to 1/N$
\begin{equation}
P_N(x) \to A_N\, N^{-N/2}\, (1-Nx)^{(N^2+N-4)/2} 
\label{limup}
\end{equation}

In contrast, deriving the behavior of $P_N(x)$ as $x\to 0$ is slightly more tricky. To derive this,
we first use the following identity of the hypergeometric function~\cite{GR}
\begin{equation}
_2F_1(\alpha,\beta,\gamma,z)= (1-z)^{-\beta}\, _2F_1(\beta, \gamma-\alpha,\gamma, z/(z-1))
\label{hyperiden1}
\end{equation}
to rewrite
\begin{equation}
P_N(x)= \frac{A_N}{\sqrt{x}}\, (1-Nx)^{(N^2+N-4)/2}\, [1-(N-1)x]^{-(N-1)/2}\, _2F_1\left(\frac{N-1}{2}, 
\frac{N^2-4}{2}, \frac{N^2+N-2}{2}, \frac{1-Nx}{1-(N-1)x}\right)
\label{new1}
\end{equation}
Now, in this form, it is easy to take the limit $x\to 0$. One gets, as $x\to 0$,
\begin{equation}
P_N(x) \to \frac{A_N}{\sqrt{x}}\, _2F_1\left(\frac{N-1}{2},
\frac{N^2-4}{2}, \frac{N^2+N-2}{2}, 1\right).
\label{new2}
\end{equation}
Using further the following identity~\cite{GR}
\begin{equation}
_2F_1(\alpha,\beta,\gamma,1)= 
\frac{\Gamma(\gamma)\,\Gamma(\gamma-\alpha-\beta)}{\Gamma(\gamma-\alpha)\,\Gamma(\gamma-\beta)}
\label{hyperiden2}
\end{equation}
and the expression for $A_N$ in Eq. (\ref{AN}) we get, as $x\to 0$
\begin{equation}
P_N(x) \to \left[\frac{\sqrt{\pi}\,
\Gamma(N)\,\Gamma(N^2/2)}{2^{N-1}\,\Gamma^2(N/2)\,\Gamma((N-1)/2)}\right]\, x^{-1/2}.
\label{lim0}
\end{equation}

\section{Asymptotic behavior of $\mu_1(N)$ for large $N$ for the real case}

In this appendix we derive the asymptotic behavior for large $N$ of $\mu_1(N)$ for the real case
given in Eq. (\ref{ravg}). We first use the following integral representation of the hypergeometric 
function~\cite{GR}
\beq
_2F_1(\alpha,\beta,\gamma,z)= \frac{1}{B(\beta,\gamma-\beta)}\,\int_0^1 t^{\beta-1}\, (1-t)^{\gamma-\beta-1}\, 
(1-tz)^{-\alpha}\, dt,
\label{intrep1}
\eeq
where $B(x,y)=\Gamma(x)\Gamma(y)/\Gamma(x+y)$ is the standard Beta function. Using this representation, we
can express $\mu_1(N)$ in Eq. (\ref{ravg}) as an integral
\beq
\mu_1(N)= \frac{2^{4-N}\Gamma(N)}{N^2 \Gamma^2(N/2)}\,\int_0^1 t^{1/2}\, (1-t)^{N/2}\, [1+(N-1)t]^{-3}\, dt,
\label{intrep2}
\eeq
which is still exact for all $N$. Next we consider the integral above, rescale $t=x/N$ and then take the
large $N$ limit as follows,
\begin{eqnarray}
\int_0^{1} t^{1/2}\, (1-t)^{N/2}\, [1+(N-1)t]^{-3}\, dt &=&  \frac{1}{N^{3/2}}\int_0^{N}  
x^{1/2}\, 
(1-x/N)^{N/2}\, \left[1+ \frac{N-1}{N} x\right]^{-3}\, dx \nonumber \\
&\approx & \frac{1}{N^{3/2}} \int_0^{\infty} \frac{x^{1/2}e^{-x/2}}{(1+x)^3}\, dx.
\label{as1}
\end{eqnarray}
Also, by Stirling's formula, $\Gamma(N)\approx \sqrt{2\pi}\, N^{N-1/2}\, e^{-N}$ for large $N$.
Using these results in Eq. (\ref{intrep2}) we get, to leading order for large $N$,
\beq
\mu_1(N) \approx \frac{c}{N^3}
\label{as2}
\eeq
where the prefactor $c$ is given by the expression
\beq
c= \frac{4\sqrt{2}}{\sqrt{\pi}}\,\int_0^{\infty} \frac{x^{1/2}e^{-x/2}}{(1+x)^3}\, dx = 
2\left[1-\sqrt{\frac{\pi e}{2}}\,{\rm erfc}(1/\sqrt{2})\right]=0.688641\ldots
\label{as3}
\eeq
where ${\rm erfc}(x)= \frac{2}{\sqrt{\pi}}\,\int_x^{\infty} e^{-u^2}\, du$ is the complementary
error function.

\end{document}